\documentclass[aps,pra,showpacs,superscriptaddress,preprint]{revtex4}
\usepackage{amsmath}
\usepackage{epsf}
\usepackage{epsfig}

\begin{document}

\title{Path Integral Solution of PT-/non-PT-Symmetric and non-Hermitian
Morse Potential}
\author{Nalan Kandirmaz$^a$ and Ramazan Sever$^b$ \thanks{
Corresponding Author: sever@metu.edu.tr}}
\address{$^a$ Mersin University, Department of Physics,Mersin\\
$^b$ Middle East Technical University, Department of Physics,Ankara}
\date{\today}
\begin{abstract}

Path integral solutions are obtained for the the
PT-/non-PT-Symmetric and non-Hermitian Morse Potential. Energy
eigenvalues and the corresponding wave functions are obtained.

\end{abstract}

\maketitle

\section{Introduction}

The concept of PT-symmetry has received much interest in recent
years in one dimensional solutions of some quantum mechanical
problems. In the standard axiom of quantum mechanics, to have a real
energy spectrum, Hamiltonian must be hermitian: $H=H^{\dag }$. In
the PT- symmetric quantum mechanics which is an alternative to
standard axiom case, Hamiltonian has real spectrum although it is
not Hermitian. When PT-symmetry is not spontaneously broken,
PT-symmetric and non hermitian complex potentials have a real energy
spectrum $[1]$. If \ any potential under the transformations
$x\rightarrow -x $ (or $x\rightarrow \xi -x$) and $i\rightarrow -i$,
satisfies $V(-x)=V^{\ast }(-x)$, it is said to be PT-symmetric.
PT-symmetric and non-Hermitian Hamiltonians having real and/or
complex eigenvalues are calculated energy spectrum and corresponding
wave functions numerically and analytically $\left[ 4,5\right] $.

In this work, Feynman's path integral method is used in order to get
energy spectrum and wave functions of PT-/non-PT-Symmetric and
non-Hermitian Morse Potential $\left[ 14,15\right] $ which is
exactly solvable $\left[ 2,10\right] $. The potentials are solved
reducing to quadratic forms with a parametric time transformation
and a canonical point transformation. The Morse potential is used to
describe interaction of the atoms in the diatomic molecules. $\left[
17\right] $. The generalized Morse potential is

\begin{equation}
V(x)=V_{1}e^{-2\alpha (r-r_{0})}-V_{2}e^{-\alpha (r-r_{0})}.
\label{EQ1}
\end{equation}%
where $r$ is the internuclear distance between the two atoms,
$r_{0}$ is location of the potential minimum and $V_{1},V_{2}$
parameters are functions determined well depth.

The paper is organized as follows: In section II, we introduce the
calculation of the energy eigenvalues and the corresponding wave
functions of Generalized Morse Potential by using Path integral
method. In section III and IV, solutions of PT-/non-PT-symmetric and
non Hermitian forms of the generalized Morse potentials are
presented by using Path integral method. We summarize the
conclusions in section V.

\section{Generalized Morse Potential}

We use path integral technique developed by Duru and Kleinert [2] to
calculate the energy eigenvalues and the corresponding wave function
of PT-/non-PT-Symmetric and non-Hermitian generalized Morse
Potential. The kernel is defined by the usual phase space path
integral in cartesian coordinates:
\begin{equation}
K(x_{b},t_{b};x_{a},t_{a})=\int \frac{DxDp}{2\pi }exp\{i\int dt[p\dot{x}-%
\frac{p^{2}}{2m}-V(x)]\}  \label{EQ2}
\end{equation}%
$(\hbar =1)$. This is the probability amplitude of a particle
traveling from a position $x_{a}$ at time $t_{a}$ to $x_{b}$ at time
$t_{b}$. The time interval can be divided into n-equal parts. Thus
one can get
\begin{equation}
t_{j}-t_{j-1}=t_{b}-t_{a}=T~\ \ \ \ \ \ \ \ \ \ j=1,2,3...N  \label{EQ3}
\end{equation}%
and

\begin{equation}
x_{j}=x(t_{j}),~~x_{0}=x_{a},~~x_{N}=x_{b}.  \label{EQ4}
\end{equation}%
The kernel can be rewritten as the limiting case of the usual time
graded form

\begin{equation}
K(x_{b},T;x_{a},0)=\int\limits_{-\infty }^{\infty
}\prod\limits_{i=1}^{n}dx_{i}\prod\limits_{j=1}^{n+1}\frac{dp_{i}}{2\pi }%
exp\{i\sum\limits_{j=1}^{n+1}[p_{i}(x_{i}-x_{i-1})-\frac{p_{i}^{2}}{2m}%
-V(x_{i})]\}  \label{EQ5}
\end{equation}%
or

\begin{equation}
K(x_{b},T;x_{a},0)=\int\limits_{-\infty }^{\infty
}\prod\limits_{j=1}^{n}dx_{j}\prod\limits_{j=1}^{n+1}\frac{dp_{j}}{2\pi }%
exp\{i\sum\limits_{j=1}^{n+1}[p_{j}(x_{j}-x_{j+1})-\frac{p_{i}^{2}}{2m}%
-V(x_{j})]\}.  \label{EQ6}
\end{equation}%
These forms are the same, since in the application of the point
transformation, $dp_{0}$ integration comes in the use of Eq. (6)
while a $dp_{n+1}$ momentum integration comes in  the Eq. (5) case.
We apply the point transformation by using Eq. (5). To have a
solvable path integral form for the generalized Morse potential, we
define

\begin{equation}
e^{-\alpha x}=u^{2}~\ \ \ \ \ \ \ \ p_{x}=-\frac{\alpha u}{2}p_{u}.
\label{EQ7}
\end{equation}%
Thus, the contribution to Jacobien becomes\

\begin{equation}
\frac{DxDp}{2\pi }=-\frac{\alpha }{2}u_{b}\frac{DuDp_{u}}{2\pi }  \label{EQ8}
\end{equation}%
and the transformed kernel takes

\begin{equation}
K(x_{b},T;x_{a},0)=-\frac{\alpha }{2}u_{b}\int \frac{DuDp_{u}}{2\pi }%
exp[i\int dt(p_{u}\dot{u}-\frac{u^{2}}{2(4m/\alpha ^{2})}%
p_{u}^{2}-V_{1}u^{4}+V_{2}u^{2})].  \label{EQ9}
\end{equation}%
To eliminate the $u^{2}$ term in the kinetic energy part, we
introduce a new time parameter $s$ $\left[ 2,10\right] $, such that

\begin{equation}
\frac{dt}{ds}=\frac{1}{u^{2}}~~or~~t=\int \frac{ds^{\prime }}{%
u^{2}(s^{\prime })}.  \label{EQ10}
\end{equation}%
Using Fourier expression of $\delta -$ function and $S=s_{b}-s_{a}$,
parametric time definition can be written as

\begin{equation}
1=\int dS\frac{1}{u_{b}^{2}}\delta (T-\int \frac{ds}{u^{2}})=\int dS\int
\frac{dE}{2\pi }\frac{1}{u_{b}^{2}}exp[i(ET-\int ds\frac{E}{u^{2}})]
\label{EQ11}
\end{equation}%
Therefore Eq. (9) becomes

\begin{equation}
K(x_{b},T;x_{a},0)=\frac{\alpha }{2}\int\limits_{-\infty }^{\infty }\frac{dE%
}{2\pi }e^{iET}\int\limits_{0}^{\infty }dSe^{iV_{2}S}\int (\frac{1}{u_{b}})%
\frac{DuDp_{u}}{2\pi }exp[i\int\limits_{0}^{S}ds(p_{u}\dot{u}-\frac{p_{u}^{2}%
}{2M}-\frac{1}{2}M\omega ^{2}u^{2}-\frac{E}{u^{2}})]  \label{EQ12}
\end{equation}%
where

\begin{equation}
M=\frac{4m}{\alpha ^{2}},~~\omega =\sqrt{\frac{V_{1}}{M}}.  \label{EQ13}
\end{equation}%
Eq. (12) can be rewritten as

\begin{equation}
K(x_{b},T;x_{a},0)=\frac{\alpha }{2}\int\limits_{-\infty }^{\infty }\frac{dE%
}{2\pi }e^{iET}\int\limits_{0}^{\infty }dSe^{iV_{2}S}K(u_{b},S;u_{a},0).
\label{EQ14}
\end{equation}%
If we integrate over $Dp_{u}$, we get

\begin{equation}
K(u_{b},S;u_{a},0)=\int (\frac{1}{u_{b}})Duexp[i\int\limits_{0}^{S}ds(\frac{1%
}{2}M\dot{u^{2}}-\frac{1}{2}M\omega ^{2}u^{2}-\frac{E}{u^{2}})].
\label{EQ15}
\end{equation}
The factor $\frac{1}{u_{b}}$ here is the result of $\delta
$-function normalization and transformation in Eq. (8). It appears
in the use of Eq. (6). To symmetrize it, we write as

\begin{equation}
\frac{1}{u_{b}}=\frac{1}{(u_{a}u_{b})^{1/2}}exp(-\frac{1}{2}ln\frac{u_{b}}{%
u_{a}})=\frac{1}{\sqrt{u_{a}u_{b}}}exp(-\frac{1}{2}\int\limits_{0}^{S}ds%
\frac{\dot{u}}{u})=\frac{1}{\sqrt{u_{a}u_{b}}}exp[i\int\limits_{0}^{S}ds%
\frac{i\dot{u}}{2u}].  \label{EQ16}
\end{equation}%
Therefore Eq. (14) can be written

\[
K(u_{b},S;u_{a},0)=\frac{1}{\sqrt{u_{a}u_{b}}}\int \frac{DuDp_{u}}{2\pi }
\]%
\begin{equation}
\times exp[i\int\limits_{0}^{S}ds(p_{u}\dot{u}-\frac{p_{u}^{2}}{2M}-\frac{1}{%
2}M\omega ^{2}u^{2}-\frac{2ME-1/4}{2Mu^{2}}+\frac{ip_{u}}{2Mu})].
\label{EQ17}
\end{equation}%
Symmetrizing the factor $\frac{1}{u_{a}}$ in the same way, we get

\begin{equation}
\frac{1}{u_{a}}=\frac{1}{\sqrt{u_{a}u_{b}}}exp[-i\int\limits_{0}^{S}ds\frac{i%
\dot{u}}{2u}]  \label{EQ18}
\end{equation}%
and kernel can be written as
\[
K(u_{b},S;u_{a},0)=\frac{1}{\sqrt{u_{a}u_{b}}}\int \frac{DuDp_{u}}{2\pi }
\]%

\begin{equation}
\times exp[i\int\limits_{0}^{S}ds(p_{u}\dot{u}-\frac{p_{u}^{2}}{2M}-\frac{1}{%
2}M\omega ^{2}u^{2}-\frac{2ME-1/4}{2Mu^{2}}-\frac{ip_{u}}{2Mu})].
\label{EQ19}
\end{equation}%
Quantum mechanical contribution to the kernel in Eq. (17) is $\frac{1}{8Mu^{2}}+%
\frac{ip_{u}}{2Mu}.$ While This term in Eq. (19) is due to
Jacobien's symmetry. Since the kernels in Eqs. (5) and (6) are
equivalent. Eq. (17) and Eq. (19) must be equivalent. Thus using
geometric average, we obtain

\begin{equation}
\overline{K}(u_{b},S;u_{a},0)=\frac{1}{\sqrt{u_{a}u_{b}}}\int \frac{DuDp_{u}%
}{2\pi }exp\{i\int\limits_{0}^{S}ds[p_{u}\dot{u}-\frac{p_{u}^{2}}{2M}-\frac{1%
}{2}M\omega ^{2}u^{2}-\frac{2ME-1/4}{2Mu^{2}}]\}.  \label{EQ20}
\end{equation}%
It has an effective Hamiltonian as seen

\begin{equation}
H_{eff}=\frac{p_{u}^{2}}{2M}+\frac{1}{2}M\omega ^{2}+\frac{2ME-1/4}{2Mu^{2}}.
\label{EQ21}
\end{equation}%
So the problem

\begin{equation}
V\left( u(s)\right) =\frac{1}{2}M\omega ^{2}u^{2}+\frac{2ME-1/4}{2Mu^{2}}
\label{EQ22}
\end{equation}%
is taken as the new potential of the moving particle. Thus we see
that the solution is reduced to harmonic oscillator case in polar
coordinates $\left[ 8\right] $. $\overline{K}(u_{b},S;u_{a},0)$ can
be obtained as

\begin{equation}
\overline{K}(u_{b},S;u_{a},0)=\frac{M\omega \sqrt{u_{a}u_{b}}}{i\sin \omega S%
}\exp \left[ \frac{iM\omega }{2}\left( u_{a}^{2}+u_{b}^{2}\right) \cot
\omega S\right] I_{\sqrt{2ME}}\left( \frac{M\omega u_{a}u_{b}}{i\sin \omega S%
}\right) .  \label{EQ23}
\end{equation}%
The energy eigenvalues and wave functions for generalized Morse potential
can be calculated by using Eq. (23). For this, we use the Hille-Hardy formula $%
\left[ 21\right] $

\begin{equation}
\frac{te^{-\alpha /2}}{1-t}\exp \left[ -\frac{1}{2}(x+y)\frac{1+t}{1-t}%
\right] I_{\alpha }\left( \frac{2\sqrt{xyt}}{1-t}\right) =%
\mathop{\displaystyle\sum}%
\limits_{n=0}^{\infty }\frac{t^{n}n!e^{-\frac{1}{2}(x+y)}}{\Gamma (n+\alpha
+1)}(xy)^{\alpha /2}L_{n}^{(\alpha )}(x)L_{n}^{(\alpha )}(y).  \label{EQ24}
\end{equation}%
Substituting $t=e^{-2i\omega S},~x=M\omega u_{a}^{2}$ \ and $y=M\omega
u_{b}^{2}$ in Eq. (23), we can write Eq. (23) as

\begin{equation}
\overline{K}(u_{b},S;u_{a},0)=%
\mathop{\displaystyle\sum}%
\limits_{n=0}^{\infty }e^{i\varepsilon _{n}S}\psi _{n}(u_{b})\psi _{n}^{\ast
}(u_{a})  \label{EQ25}
\end{equation}%
where the energy $\varepsilon _{n}$ is given by

\begin{equation}
\varepsilon _{n}=\omega \left( 2n+1+\sqrt{2ME}\right) .  \label{EQ26}
\end{equation}%
From Eq. (24) wave functions are obtained as

\begin{equation}
\psi _{n}(u)=\sqrt{\frac{2M\omega n!\Gamma (n+2s+\frac{1}{2})}{un!}}\left(
M\omega u^{2}\right) ^{s+\frac{1}{2}}\exp \left( -M\omega u^{2}\right)
L_{n}^{2s+\frac{1}{2}}\left( M\omega u^{2}\right) .  \label{EQ27}
\end{equation}%
Here $s=\frac{1}{4}+\frac{1}{2}\sqrt{2mE}>0,~~~n=0,1,2...$ and $L_{n}^{2s+%
\frac{1}{2}}\left( M\omega u^{2}\right) $ is associated Laguerre
polynomials. Therefore we calculate the energy dependent Greens
function for the generalized Morse potential
\[
G(u_{b},u_{a};E)=-\frac{\alpha }{2\sqrt{u_{a}u_{b}}}%
\displaystyle\int %
\limits_{-\infty }^{\infty }\frac{dE}{2\pi }e^{iET}%
\mathop{\displaystyle\sum}%
\limits_{n=0}^{\infty }%
\displaystyle\int %
\limits_{0}^{\infty }dSe^{(iV_{2}-\varepsilon _{n})S}\psi _{n}(u_{b})\psi
_{n}^{\ast }(u_{a})
\]%

\begin{equation}
=%
\mathop{\displaystyle\sum}%
\limits_{n=0}^{\infty }\exp \left\{ -i\left[ -V_{2}\left( 1-\frac{2\omega }{%
V_{2}}(n+\frac{1}{2})\right) ^{2}\right] T\right\} \phi _{n}(u_{b})\phi
_{n}^{\ast }(u_{a}).  \label{EQ28}
\end{equation}%
Integrating over $dS$ and $dE,$ we can get energy eigenvalues

\begin{equation}
E_{n}=-V_{2}\left[ 1-\frac{2\omega }{V_{2}}(n+\frac{1}{2})\right]
^{2},~~~n=0,1,2...<\frac{V_{2}}{\omega }-\frac{1}{2}  \label{EQ29}
\end{equation}%
\newline
and normalized wave functions are%
\begin{equation}
\phi _{n}(u)=\sqrt{\frac{2\alpha (s-\frac{1}{4})M\omega n!}{\Gamma (n+2s+%
\frac{1}{2})}}\left( M\omega u^{2}\right) ^{s+\frac{1}{2}}\exp \left(
-M\omega u^{2}\right) L_{n}^{2s+\frac{1}{2}}\left( M\omega u^{2}\right)
\label{EQ30}
\end{equation}%
Therefore, PT-symmetric and Hermitian generalized Morse potential
have these wave functions and energy eigenvalues.

\section{\protect\normalsize PT-symmetric and non-Hermitian Morse Potential
Case}

{\normalsize If $V_{1}$ and$~V_{2}$ are real and $\alpha =i\alpha $ then the
Morse potential has the form }

{\normalsize
\begin{equation}
V(x)=V_{1}e^{-2i\alpha x}-V_{2}e^{-i\alpha x}.  \label{EQ31}
\end{equation}%
We can get wave functions and energy eigenvalues following the same
steps. This time, we use a coordinate transformation $\
u(t)=e^{-i\alpha x}$  and a new time parameter }$s$ defined as

{\normalsize
\begin{equation}
\frac{dt}{ds}=-\frac{1}{u^{2}}.  \label{EQ32}
\end{equation}%
\ Thus the kernel becomes }

\begin{eqnarray}
K(x_{b},T;x_{a},0) &=&-\frac{i\alpha }{2}\int\limits_{-\infty }^{\infty }%
\frac{dE}{2\pi }e^{iET}\int\limits_{0}^{\infty }dSe^{-iV_{2}S}\int (\frac{1}{%
u_{b}})\frac{DuDp_{u}}{2\pi }  \nonumber \\
&&\times exp[i\int\limits_{0}^{S}ds(p_{u}\dot{u}-\frac{p_{u}^{2}}{2M}-\frac{1%
}{2}M\omega ^{2}u^{2}+\frac{E}{u^{2}})]  \label{EQ33}
\end{eqnarray}%
where

\begin{equation}
M=\frac{4m}{\alpha ^{2}},~~\omega =\sqrt{-\frac{V_{1}}{M}}.  \label{EQ34}
\end{equation}%
{\normalsize Frequency is defined as $\omega =\sqrt{\frac{%
-V_{1}}{M}}$ and energy is $V_{2}$ . }Applying the symmetrization of
Jacobien in the same way, kernel takes

\begin{equation}
\overline{K}(u_{b},S;u_{a},0)=\frac{1}{\sqrt{u_{a}u_{b}}}\int \frac{DuDp_{u}%
}{2\pi }exp\{i\int\limits_{0}^{S}ds[p_{u}\dot{u}-\frac{p_{u}^{2}}{2M}-\frac{1%
}{2}M\omega ^{2}u^{2}-\frac{\left( \sqrt{-2ME}\right) ^{2}-1/4}{2Mu^{2}}]\}.
\label{EQ35}
\end{equation}%
So, effective Hamiltonian of the system is written as%

\begin{equation}
H_{eff}=\frac{p_{u}^{2}}{2M}+\frac{1}{2}M\omega ^{2}+\frac{\left( \sqrt{-2ME}%
\right) ^{2}-1/4}{2Mu^{2}}.  \label{EQ36}
\end{equation}%
Applying the same procedure, kernel can be obtained as%

\begin{equation}
\overline{K}(u_{b},S;u_{a},0)=\frac{M\omega \sqrt{u_{a}u_{b}}}{i\sin \omega S%
}\exp \left[ \frac{iM\omega }{2}\left( u_{a}^{2}+u_{b}^{2}\right) \cot
\omega S\right] I_{\sqrt{-2ME}}\left( \frac{M\omega u_{a}u_{b}}{i\sin \omega
S}\right) .  \label{EQ37}
\end{equation}%
Therefore, Greens function becomes%
\[
G(u_{b},u_{a};E)=-\frac{i\alpha }{2\sqrt{u_{a}u_{b}}}%
\displaystyle\int %
\limits_{-\infty }^{\infty }\frac{dE}{2\pi }e^{iET}%
\mathop{\displaystyle\sum}%
\limits_{n=0}^{\infty }%
\displaystyle\int %
\limits_{0}^{\infty }dSe^{(-iV_{2}-\varepsilon _{n})S}\psi _{n}(u_{b})\psi
_{n}^{\ast }(u_{a})
\]%

\begin{equation}
=%
\mathop{\displaystyle\sum}%
\limits_{n=0}^{\infty }\exp \left\{ -i\left[ V_{2}\left( 1+\frac{2\omega }{%
V_{2}}(n+\frac{1}{2})\right) ^{2}\right] T\right\} \phi _{n}(u_{b})\phi
_{n}^{\ast }(u_{a}).  \label{EQ38}
\end{equation}%
Energy eigenvalues are

\begin{equation}
E_{n}=-V_{2}\left[ 1+\frac{2\omega }{V_{2}}(n+\frac{1}{2})\right]
^{2},~~~n=0,1,2...<\frac{V_{2}}{\omega }+\frac{1}{2}  \label{EQ39}
\end{equation}%
and normalized wave functions become

\begin{equation}
\phi _{n}(u)=\sqrt{\frac{2\alpha (s-\frac{1}{4})M\omega n!}{\Gamma (n+2s+%
\frac{1}{2})}}\left( M\omega u^{2}\right) ^{s+\frac{1}{2}}\exp \left(
-M\omega u^{2}\right) L_{n}^{2s+\frac{1}{2}}\left( M\omega u^{2}\right) ,
\label{EQ40}
\end{equation}%
where $s=\frac{1}{4}+\frac{1}{2}\sqrt{-2mE},~~~n=0,1,2...$

\bigskip

\section{\protect\normalsize Non-PT-symmetric and non-Hermitian Morse
Potential case}

If we take $V_{1}=\left( A+iB\right) ^{2},~V_{2}=\left( 2C+1\right)
\left( A+iB\right) $ and $\alpha =1$,  the Morse potential takes
\begin{equation}
V(x)=\left( A+iB\right) ^{2}e^{-2x}-\left( 2C+1\right) \left( A+iB\right)
e^{-x}.  \label{EQ41}
\end{equation}%
Here $A,B,C$ are arbitrary parameters. This potential is non-PT
symmetric and non-Hermitian, but has real spectra. If $V_{1}$ is
real $~V_{2}=A+iB$ and $\alpha =i\alpha $ the Morse potential has
the form

\begin{equation}
V(x)=V_{1}e^{-2i\alpha x}-\left( A+iB\right) e^{-i\alpha x}.  \label{EQ42}
\end{equation}%
We can derive wave functions and energy eigenvalues in the same way.
This time, we use a coordinate transformation $\ u(t)=e^{-i\alpha
x}$ and \ define a new time parameter$,${\normalsize \ then the
kernel becomes }
\[
K(x_{b},T;x_{a},0)=-\frac{i\alpha }{2}\int\limits_{-\infty }^{\infty }\frac{%
dE}{2\pi }e^{iET}\int\limits_{0}^{\infty }dSe^{-i(A+iB)S}\int (\frac{1}{u_{b}%
})\frac{DuDp_{u}}{2\pi }exp[i\int\limits_{0}^{S}ds(p_{u}\dot{u}-\frac{%
p_{u}^{2}}{2M}-\frac{1}{2}M\omega ^{2}u^{2}+\frac{E}{u^{2}})]
\]%

\begin{equation}
\times exp[i\int\limits_{0}^{S}ds(p_{u}\dot{u}-\frac{p_{u}^{2}}{2M}-\frac{1}{%
2}M\omega ^{2}u^{2}+\frac{E}{u^{2}})]  \label{EQ43}
\end{equation}%
where

\begin{equation}
M=\frac{4m}{\alpha ^{2}},~~\omega =\sqrt{-\frac{V_{1}}{M}}  \label{EQ44}
\end{equation}%
Frequency is defined as $\omega =\sqrt{\frac{-V_{1}}{M}}$ and energy
is $(A+iB).${\normalsize \ }Applying the symmetrization of Jacobien
in the same way, kernel becomes
\begin{equation}
\overline{K}(u_{b},S;u_{a},0)=\frac{1}{\sqrt{u_{a}u_{b}}}\int \frac{DuDp_{u}%
}{2\pi }exp\{i\int\limits_{0}^{S}ds[p_{u}\dot{u}-\frac{p_{u}^{2}}{2M}-\frac{1%
}{2}M\omega ^{2}u^{2}-\frac{\left( \sqrt{-2ME}\right) ^{2}-1/4}{2Mu^{2}}]\}.
\label{EQ45}
\end{equation}%
So, effective Hamiltonian of the system is written as%

\begin{equation}
H_{eff}=\frac{p_{u}^{2}}{2M}+\frac{1}{2}M\omega ^{2}+\frac{\left( \sqrt{-2ME}%
\right) ^{2}-1/4}{2Mu^{2}}.  \label{EQ46}
\end{equation}%
Applying the same procedure, kernel can be obtained as%

\begin{equation}
\overline{K}(u_{b},S;u_{a},0)=\frac{M\omega \sqrt{u_{a}u_{b}}}{i\sin \omega S%
}\exp \left[ \frac{iM\omega }{2}\left( u_{a}^{2}+u_{b}^{2}\right) \cot
\omega S\right] I_{\sqrt{-2ME}}\left( \frac{M\omega u_{a}u_{b}}{i\sin \omega
S}\right) .  \label{EQ47}
\end{equation}%
Therefore, Greens function becomes

\begin{equation}
G(u_{b},u_{a};E)=-\frac{i\alpha }{2\sqrt{u_{a}u_{b}}}%
\displaystyle\int %
\limits_{-\infty }^{\infty }\frac{dE}{2\pi }e^{iET}%
\mathop{\displaystyle\sum}%
\limits_{n=0}^{\infty }%
\displaystyle\int %
\limits_{0}^{\infty }dSe^{[-i(A+iB)-\varepsilon _{n}]S}\psi _{n}(u_{b})\psi
_{n}^{\ast }(u_{a}).  \label{EQ48}
\end{equation}%
energy eigenvalues are%

\begin{equation}
E_{n}=-\left( A+iB\right) \left[ 1+\frac{2\omega }{A+iB}(n+\frac{1}{2})%
\right] ^{2},~~~n=0,1,2...<\frac{A+iB}{\omega }+\frac{1}{2}  \label{EQ49}
\end{equation}%
and normalized wave functions become

\begin{equation}
\phi _{n}(u)=\sqrt{\frac{2\alpha (s-\frac{1}{4})M\omega n!}{\Gamma (n+2s+%
\frac{1}{2})}}\left( M\omega u^{2}\right) ^{s+\frac{1}{2}}\exp \left(
-M\omega u^{2}\right) L_{n}^{2s+\frac{1}{2}}\left( M\omega u^{2}\right)
\label{EQ50}
\end{equation}%
where $s=\frac{1}{4}+\frac{1}{2}\sqrt{-2mE},~~~n=0,1,2...$ \bigskip

\section{Conclusion}

The energy eigenvalues and the corresponding wave functions for
PT-/non-PT-Symmetric and non-Hermitian generalized Morse Potential
are obtained by using Path integral method. The real energy spectra
of the PT-/non-PT-symmetric and non-Hermitian forms of potential
have been obtained by restricting the potential parameters. The
Hamiltonian of system transformed to the form of the two oscillators
case with the frequency $\omega $ with a proper time parameter $s$.
The approach can also be applied to the other PT-/non-PT-Symmetric
and non-Hermitian potentials for which the potential problem can be
transformed to the oscillator case.

\section{Acknowledgements}

This research was partially supported by the Scientific and
Technological Research Council of Turkey.


\begin{thebibliography}{99}

\bibitem{Bender}
C.M. Bender, Reports on Progress In Physics 70 (6), 947-1018
(2007); C. M. Bender and D.W.Darg, Journal of Mathematical Physics
48 (4), 042703 (2007); C. M. Bender and S.Boettcher, Pys. Lett.
80, 5243(1998); C. M. Bender, S. Boetcher, and P.N. Meisenger, J.
Math. Phys. 40, 2201 (1999); C. M. Bender, G. V. Dunne, and P. N.
Meisenger, Phys. Lett.A, 252, 272 (1999).

\bibitem{Duru}
I.H. Duru, Phys. Rev. D, 28, 2689(1983).,I.H. Duru and H.
Kleinert, Phys. Lett. {\bf B84}, 185(1979); Fortschr. Phys. {\bf
30}, 401(1982); I.H. Duru, Phys.Rev. D, 30, 2121(1984); I.H. Duru,
Phys. Lett A, 119,163(1986).

\bibitem{Feynman}
R. Feynmann and A. Hibbs: Quantum Mechanics and Path Integrals
McGraw Hill, New York, (1965).

\bibitem{Ahmed}
Z.Ahmed, Phys. Lett. {\bf A282}, 343 (2001).

\bibitem{Bagchi}
B. Bagchi, C. Quesne, Phys. Lett. {\bf A273}, 285 (2000).

\bibitem{Goldstein}
H. Goldstein, Classical Mechanics (Addison-Wesley, Reading, MA
(1980).

\bibitem{Kustaanheimo}
P. Kustaanheimo and E. Stiefel, J. Reine Angew.
Math. {\bf 218}, 204 (1965).

\bibitem{Inomata}
D. Peak and A. Inomata, J. Math. Phys. 10, 1422 (1969).

\bibitem{Toyoda}
T. Toyoda and S. Wakayama, Phys. Rev. {\bf A59}, 1021 (1999).

\bibitem{Pak}
N.K. Pak and I. Sokmen, Phys. Lett. 100A, 327(1984); N.K. Pak and
I. Sokmen, Phys. Rev. A, 30, 1629(1984); N.K. Pak and I. Sokmen,
Phys. Lett. A, 103, 298 (1984).

\bibitem{Jia}
C.S.Jia,Y.Sun and Y. Li, Phys. Lett.A, 305,231 (2002).

\bibitem{Unal}
N. Unal, Phys. Can. J. Phys. 80, 875 (2002).

\bibitem{Cannata}
F. Cannata, G. Junker and J. Trost, Phys. Lett A246, 219 (1998).

\bibitem{Yesiltas}
O. Yesiltas, M. Simsek, R. Sever, C. Tezcan, Phys. Scripta, T67,
472 (2003).

\bibitem{Berkdemir}
C. Berkdemir, A. Berkdemir and R.Sever, Phys. Rev. C72, 027001
(2005).

\bibitem{Znojil}
M. Znojil, Phys. Lett. A264, 108 (1999).

\bibitem{Morse}
P. M. Morse. Phys. Rev. 34, 57 (1929).

\bibitem{Levine}
R. D. Levine and C.E. Wulfman. Chem. Phys. Lett. 60, 372 (1979).

\bibitem{Sokmen}
I. Sokmen, Phys. Lett. A115, 249 (1986).

\bibitem{Kibler}
M. Kibler, T. Negadi, Theor. Chim. Acta 66, 31 (1984).

\bibitem{Gradshteyn}
I.S. Gradshteyn and I.M. Ryzhic:Table of Integrals, Series, and
Products, Academic Press, New York, (1980).

\end{thebibliography}
\end{document}